\documentclass[aps,prb,showpacs,groupedaddress,twocolumn]{revtex4}
\usepackage{amsmath,amssymb}

\usepackage{graphicx} % Include figure files
\usepackage{dcolumn} % Align table columns on decimal point
\usepackage{bm} % bold math

\begin{document}
\draft
\title{\bf Opposite spin accumulations on the transverse edges
by the confining potential}
\author{Yanxia Xing$^1$, Qing-feng Sun$^{1,*}$,
Liang Tang$^2$ and  JiangPing Hu$^{2,**}$ }
\address{
$^1$Beijng National Laboratory for Condensed Matter Physics and
Institute of Physics, Chinese Academy of Sciences, Bejing 100080,
P.R. China\\
$^2$Department of Physics, Purdue University, West Lafayette, Indiana 47907, USA   }

%\maketitle

\begin{abstract}
We show that the spin-orbit interaction induced by the boundary
confining potential causes  opposite spin accumulations on the
transverse edges in a zonal two-dimensional electron gas in the
presence of external longitudinal electric field.  While the bias
is reversed, the spin polarized direction  is also reversed. The
intensity of the spin accumulation is proportional to the bias
voltage. In contrast to the bulk extrinsic and intrinsic spin Hall
effects, the spin accumulation by the confining potential is
almost unaffected by impurity and survives even in strong
disorder. The result provides a new mechanism to explain the
recent experimental data.

\end{abstract}

 \maketitle

\section {Introduction}

Recently, two experimental groups observed the transverse opposite
spin accumulations near two edges of their devices  in the
presence of a longitudinal voltage
bias.\cite{experiment1,experiment2}  One experiment is on n-type
GaAs's bar with a size about $300 \mu m\times 77 \mu
m$,\cite{experiment1} and the spin accumulation is detected by
Kerr rotation spectroscopy. The other experiment is on a coplanar
p-n junction light-emitting diode device.\cite{experiment2} Under
a longitudinal bias, a circular polarization of the emitting light
on two edges is detected. The directions of the polarization are
opposite on the two edges, which suggests opposite spin
accumulation on the two edges. Moreover, when the bias is
reversed, the spin accumulations are reversed in the above two
experiments.

These experiments originally were  motivated to  measure the
predicted effects:  the extrinsic and intrinsic spin Hall effects
(SHE). The extrinsic SHE has been discovered about a few decades
ago,\cite{eshe1,eshe2} and it originates from the spin dependent
scattering that deflects the spin-up and spin-down carriers towards
the opposite edges of a sample. The intrinsic SHE is predicted first
by Murakami {\it et.al.} and Sinova {\sl et.al.} in a Luttinger
spin-orbit (SO) coupled 3D p-doped semiconductor\cite{Zhangsc} and a
Rashba SO coupled two-dimensional electron gas (2DEG)\cite{Sinova},
respectively. Recently, a number of sequent works have focused on
this interesting effect.\cite{SHE1,SHE2,SHE3,SHE4}  Nonetheless, the
intrinsic SHE still remains a controversy topic. The intrinsic spin
Hall conductivity originally was pointed out to be universal in the
clean bulk sample.\cite{Sinova} However some works have showed that
the spin-Hall conductivity depends on the SO coupling strength and
electron Fermi energy in general.\cite{SHE1} Moreover, it has been
shown recently that the impurity  plays an important role  on the
intrinsic SHE. In an infinite system, it has been shown that the
spin Hall conductivity is very sensitive to disorder.\cite{SHE2} The
effect vanishes even in a very weak disorder limit when  the vertex
correction is considered. In a finite mesoscopic ballistic system,
the SHE can survive.\cite{SHE3} The SHE and spin accumulations have
been studied in the dirty or clean finite mesoscopic samples by
using the Landauer-B\"{u}ttiker formalism and the tight-binding
Hamiltonian\cite{Dattabook}. These works show that the opposite spin
accumulations indeed can generate on the transverse two edges in the
finite system, and the SHE still presents below a critical disorder
threshold.\cite{SHE3}

Although the spin accumulation  observed in the two experiments
appears to reflect the physics of the SHE, there is still a
significant challenge to explain the observed experiment effect.
The spin accumulation contributed from the extrinsic SHE  has been
shown to have  the directions of spin polarization that are
opposite to the experimental results.\cite{SHE2} While the
intrinsic SHE is  predicted to have the same directions of spin
polarization as that observed in the experiments, it is expected
to be scaled with length and eventually    be vanished in Rashba
spin orbit coupling systems in the presence of disorder. Since the
experiments are not done in the ballistic region and the signal of
spin accumulation has little dependence of the transversal length
of the sample, it is not clear that  how the intrinsic SHE
explains all experimental results solely.

\begin{figure}%[tbp]
\includegraphics[bb=12mm 12mm 173mm 82mm, width=7cm,totalheight=3.0cm, clip=]{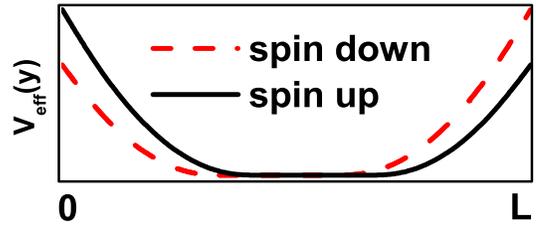}
\caption{(Color online) Schematic diagram for the transverse
effective potential $V_{eff}(y)$ in the zonal 2DEG for
the positive $+p_x$. }
\end{figure}

In this paper, we give a new mechanism  that can generate the
opposite spin accumulations in two edges under the longitudinal
voltage bias. We first show the principle of this new mechanism.
Considering an infinite zonal 2DEG with a confining potential  on
$y$-direction, which is described by the Hamiltonian $H$:
$H=\frac{p_x^2+p_y^2}{2m^*} +V(y)$, where $p_{x/y}$ is the
momentum operator and $m^*$ is the effective mass of electrons.
$V(y)$ is the confining potential energy which is constant in the
center and quickly increases  as  tending to the boundary. Based
on the relativity effect, in the presence of the internal electric
field ${\bf E} = \nabla V(y)/e$, there is  a natural spin-orbit
coupling term $V_{SO}=\frac{e \hbar}{2m^{*2}c^2}\;\sigma\cdot({\bf
E}\times{\bf p})$.\cite{book,sun} Considering that the
corresponding electric field ${\bf E} = \nabla V(y)/e$ is
perpendicular to the edge, only the element $E_y$ of
$y$-direction is non-zero and the spin-orbit coupling energy
reduces into: $V_{SO}=-\frac{\hbar}{2m^{*2}c^2}\;\sigma_z p_x
\frac{d}{dy} V(y)$. So for the spin-down ($\sigma_z=\downarrow$ or
$-1$) electrons, its effective potential
$V_{eff}(y)=V(y)+V_{SO}(y)$ is lower than the one of spin-up
electrons at the edge of $y=0$ for positive $+p_x$ (see Fig.1). On
the other hand, at the other edge of $y=L$, the effective
potential $V_{eff}$ of the spin-up electrons is lower for $+p_x$
(see Fig.1). Thus, the spin accumulations on the two transverse
edges form  when electrons occupy the positive $+p_x$ states under
the positive longitudinal  bias. When the bias is reversed, the
electrons occupy the negative $-p_x$ state and the spin
accumulation reverses its sign. Therefore it produces the same
spin accumulation as that  observed in  the
experiments.\cite{experiment1,experiment2} In the present model,
the spin accumulation   originates completely from the structure
confining potential, so it is not affected by the impurity and the
dephase. In a word, the structure confining potential can also
induce the opposite spin accumulation, which can be the origin of
the experimentally observed spin accumulation.

The paper is organized as follows: in section II, we will mention
and solve the model in detail. The results and discussions are in
section III. Finally, a brief summary is given in section IV.

\section{The model and solution}

The Hamiltonian of the zonal 2DEG can be
written as:
\begin{eqnarray}
 \label{Hamiltonian}
 H= \frac{p_x^2+p_y^2}{2m^*} + V(y)-
 \frac{\hbar}{2m^{*2}c^2}\;\sigma_z p_x \frac{d}{dy}V(y).
\end{eqnarray}
Here the first term is kinetic energy, the second term is the
potential energy, and the third term is from the spin-orbit
coupling energy due to the boundary confining potential $V(y)$ as
mentioned in the introduction.

Due to the fact $[p_x,H]=0$ and $[\sigma_z, H]=0$, $k_x$ and
$\sigma_z$ are the good quantum numbers,  the eigenstates in such
a structure can be written in the form
$\Psi(x,y)=\phi_{nk_x}(y)exp(ik_x x)$ with the dispersion relation
$E=\epsilon_{nk_x}+\hbar^2k_x^2/2m^*$. The index $n$ numbers the
different subbands   with the wave-function $\phi_{nk_x}(y)$ in
the $y$-direction and a  energy $\epsilon_{nk_x}$, which are
functions of the wave vector $k_x$. The transverse wave-function
$\phi_{nk_x}(y)$ satisfies the equation:
\begin{equation}\label{schro}
 \left[\frac{p_y^2}{2m^*} + V(y)-
 \gamma \sigma_z \frac{dV(y)}{dy}\right]
 \phi_ {nk_x}(y)=
 \epsilon_{nk_x} \phi_{nk_x}(y),
\end{equation}
where $\gamma=\frac{\hbar^2 k_x }{2m^{*2}c^2}$.
In the experiment, the length $L$ along the transverse $y$-direction
is very long [e.g. in the order of tens of micron in
Ref.(1)] while the confining potential is only
limited in several atom's layers. So we model this potential $V(y)$
as a square potential well, i.e. $V(y)=0$ for $0<y<L$ and $V(y)=V$
for others $y$. In this case,
$V_{SO}=-\gamma \sigma_z  \frac{d}{dy} V(y)$
becomes the $\delta$-function. The Hamiltonian can be solved
analytically, and the Schr\"{o}dinger equation Eq.(\ref{schro}) for
the spin-up electron reduces into:
\begin{equation}\label{up}
 [\frac{p_y^2}{2m^*} + V(y)+\gamma
 (\;\delta(y)-\delta(y-L)\;)\;]\phi(y)=\epsilon
 \phi(y) .
\end{equation}
The spin-down electronic Schr\"{o}dinger equation is same to
Eq.(\ref{up}) except $\gamma\rightarrow -\gamma$. The wave
function $\phi(y)$ in Eq.(\ref{up}) can be written as:
\begin{equation}\label{}
\phi(y)=\left\{
\begin{array}{ll}
Ae^{\beta y},&~~ y<0\\
sin(k_y y+\theta),&~~ 0<y<L\\
Be^{-\beta y},&~~ y>L
\end{array}\right. \nonumber
\end{equation}
where $k_y=\sqrt{2m^*\epsilon}/\hbar$,
$\beta=\sqrt{2m^*(V-\epsilon)}/\hbar$, and $A$, $B$, and $\theta$
are the constants to be determined by the boundary conditions. Here
the boundary conditions are
$\phi(y)|_{y=0^-/L^-}=\phi(y)|_{y=0^+/L^+}$ and
$\phi'(y)|_{y=0^+/L^+}-\phi'(y)|_{y=0^-/L^-}=\pm(2m^*\gamma/\hbar^2)\phi(y)|_{y=0/L}$,
which lead to
$$\left\{
\begin{array}{l}
sin\theta=A \\
sin(k_y L+\theta)=Be^{-\beta L} \\
k_y cos\theta-A\beta=2m^*\gamma A \\
\beta Be^{-\beta L}+k_ycos(k_y L+\theta)=2m^*\gamma Be^{-\beta L}
\end{array}\right.
$$
Solving the equation set, we get $\epsilon_{nk_x}$ and
$\phi_{nk_x}(y)$. Sequentially the spin-up electron probability
distribution $P_{nk_x,\uparrow}(y) =|\phi_{nk_x}(y)|^2$ is obtained.
The same calculation can be done for the spin-down electron. Due to
the system is universal along the $x$-direction, and the  spin
accumulation is independent of $x$, $P_{nk_x,\uparrow\downarrow}(y)$
completely describes spin distribution.

Although the square confining potential $V(y)$ model is solvable
analytically, the potential in the real system is not abrupt.  An
gradual change from bottom to top close to the interface is
expected. For this reason, we consider the real parabolic confining
potential $V(y)$  and solve the system numerically by using the
tight-binding Hamiltonian\cite{Dattabook}. In addition, we also
study the disorder effect on the spin accumulations. In the
tight-binding approximation, the Hamiltonian in Eq.(\ref{schro}),
which is related to $y$-direction, can be written as the following
discrete lattice version:
\begin{eqnarray}\label{Hamil}
 H= \sum\limits_{i,\sigma}(\epsilon_i+V_i+\sigma_z V_{SO,i})
 a_{i \sigma}^\dagger  a_{i\sigma} +
 \sum\limits_{<ij>,\sigma}t\;
 a_{i \sigma}^\dagger  a_{j\sigma} ,
\end{eqnarray}
where $\sigma=\uparrow, \downarrow$ (or $\pm 1$) is the spin index
in $z$-direction, and $ t= \hbar^2/2m^* a^2$ represents the hopping
matrix element with the lattice constant $a$. The confining
potential $V(y)$ is assumed to be parabolic: $V(y)=
V\frac{(y-6a)^2}{25a^2}$ for $a\leq y \leq 5a$, $V(y)= 0$ for $ 5a < y
\leq L-5a$, and $V(y)= V\frac{(y-L+5a)^2}{25a^2}$ for $L-5a < y \leq
L$. For a clean system, the onsite energy $\epsilon_i=0$, and
$\epsilon_i$ is
randomly distributed between [$-W/2,W/2$] for the dirty system. The
Hamiltonian in Eq.(\ref{Hamil}) can easily be solved by numerically
calculating the eigenvalues and eigenstates of the Hamiltonian
matrix.  Due to decoupling between the different spin states in the
Hamiltonian, we can solve the eigenvalues and eigen-wavefunctions
separately for the spin-up and spin-down electrons. Whereafter, the
the spin-up and spin-down electron probability distribution
$P_{nk_x,\sigma}(i)$ in the subband $n$ and the longitudinal
momentum $\hbar k_x$ is obtained straightforwardly.

While the longitudinal bias is zero, the spin accumulations $S(y)$
is zero everywhere, because of the existence of the time-reversal
invariance. On the other hand, when a bias $V_{bias}$ is added, the
spin accumulations $S(y)$ will emerge. Consider the device under the
positive bias $V_{bias}$ and at the zero temperature, the $+k_x$
states with its energy between $E_f -V_{bias}/2$ and $E_f
+V_{bias}/2$ are occupied by electrons, while the states for the
negative $-k_x$ are empty, here $E_f$ is the Fermi energy. Then,
under the small bias and taking the linear approximation, the
spatial density distribution $P_{\uparrow\downarrow}(i)$ of the
spin-up (down) electrons along $y$-direction for the unit bias is
$P_{\uparrow\downarrow}(i) = \sum_n \rho_n(E_{k_x})
P_{nk_x,\uparrow}(i)$, where $\rho_n(E_{k_x})$ is the density of
state in the subbands $n$ with $E_{k_x} = E_f - \epsilon_{nk_x}$,
and the sum is over all subbands $n$ with its cut-off energy lower
than $E_f$. The spin accumulation density and charge density in the
linear bias can be obtained as: $P_s(i) \equiv
\lim_{V_{bias}\rightarrow 0} \frac{S(ia)}{V_{bias}} = \frac{\hbar}{2}
  (P_{\uparrow}(i)-P_{\downarrow}(i) )$ and
$P_e(i) = e(P_{\uparrow}(i)+P_{\downarrow}(i) )$.

In the numerical calculation, we choose the realistic parameters as
ones in the experiment:\cite{Gui}  the electronic effective mass
$m^* =0.05 m_e$ and the Fermi energy $E_f =0.1 eV$, in which the
corresponding electron concentration is approximately
$n_{2D}=10^{12}cm^{-2}$. The energy unit is set to $1eV$, and the
length unit is $1nm$ for the analytic model, or $1a.u.$ for the
tight-banding model. The lattice constant $a$ in the tight-banding
model is around $0.196 nm$.

\section{numerical results and discussion}

\begin{figure}%[tbp]
\includegraphics[bb=11mm 11mm 197mm 146mm, width=8.5cm,totalheight=7cm,clip=]{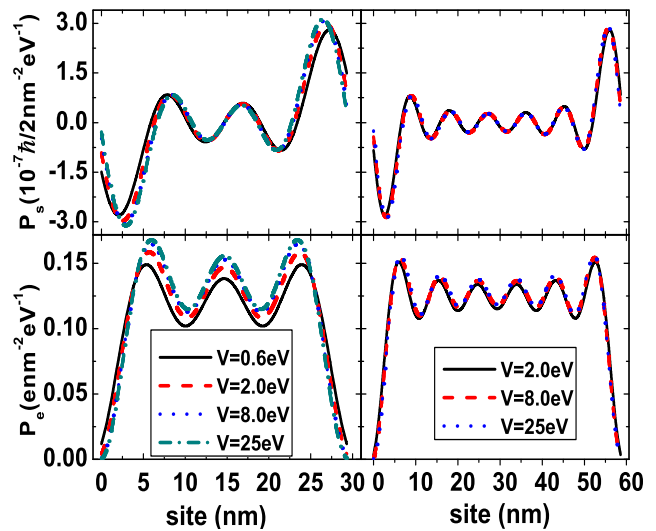}
\caption{(Color online) The transverse distribution of the spin
and charge density, $P_s(y)$ and $P_e(y)$, vs $y$ for the square
confining potential. Left panel: $L=29.3nm$, right panel:
$L=58.6nm$. }
\end{figure}
\begin{figure}%[tbp]
\includegraphics[bb=10mm 10mm 190mm 143mm, width=8.5cm,totalheight=7cm,clip=]{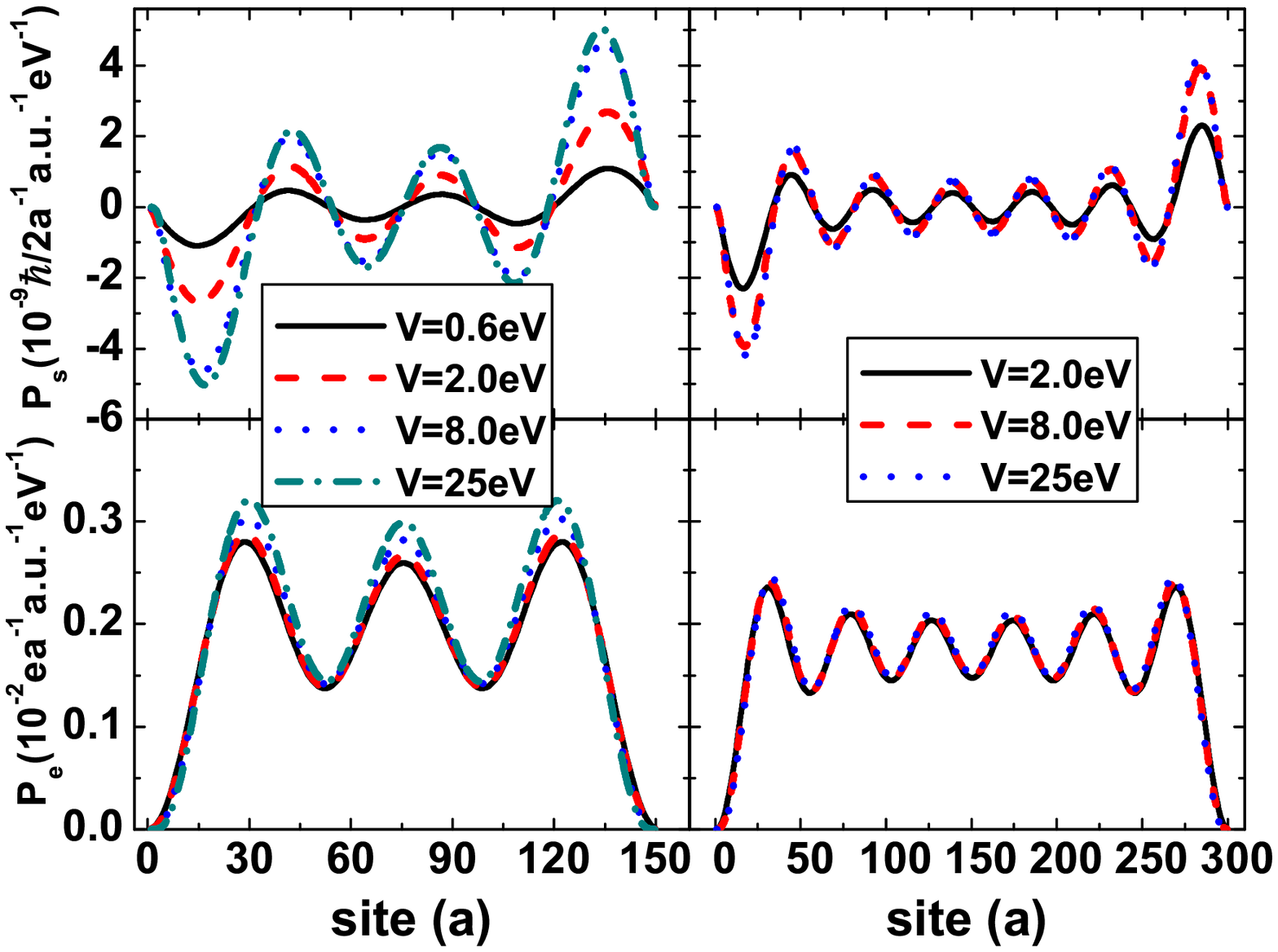}
\caption{(Color online) The transverse distribution of the spin
and charge density, $P_s(y)$ and $P_e(y)$, vs $y$ for the
parabolic confining potential. Left panel: $L=150 a$, right panel:
$L=300 a$. }
\end{figure}

First, we study spin accumulation in the clean system. In the
present system, the spin accumulation density $P_s(y)$ depends on
the transverse position $y$, and it is independent of the
longitudinal position $x$. Fig.2 and Fig.3 show the spin
accumulation density $P_s(y)$ and charge density $P_e(y)$ versus
$y$ for the case of the square and the parabolic confining
potential $V(y)$, respectively. First of all, the opposite spin
accumulations at $z$-direction indeed are generated near the two
edges regardless of the square or the parabolic potential. If the
bias is reversed, the electrons occupy the negative $-k_x$ states
instead of the positive $+k_x$ states and  the spin accumulations
also are reversed. These results are consistent with the
experimental results.\cite{experiment1,experiment2} The opposite
spin accumulations  here obviously  originates from the confining
potential as mentioned in the introduction because there is no
other interaction except for the potential $V(y)$. Second, the
spin accumulations mainly are near the two edges, and it is small
and has oscillation in the bulk. The oscillation is expected due
to the existence of Fermi surface. The period is given by
$2\pi/k_F$.  For a fixed Fermi energy (e.g. $E_f=0.1eV$), the
wider the width $L$ is, the more the subbands below $E_f$ will be.
At the mean time, the oscillation times of $P_s$ and $P_e$ are
more and the oscillation amplitude are smaller. So the bulk $P_s$
almost vanishes and $P_e$ approaches constant at large $L$. But
the spin accumulations near the edge, including the intensity and
the location, is almost independent with $L$. Third, we discuss
the characters of the spin accumulations $P_s(y)$ as a function of
the strength $V$ of the confining potential $V(y)$. The characters
are slightly different for the square and parabolic potentials.
While $V=0$, $P_s(y)=0$ for both  the square and parabolic
potentials. With $V$ increasing, $P_s(y)$ emerges. For the square
potential, $P_s(y)$ quickly arises in the beginning. Around
$V=5E_f$, $P_s(y)$ has reached saturated value. Thereafter the
value almost does not change with further increasing $V$ (see
Fig.2a,b). For the parabolic potential, $P_s(y)$ increases
slowly. Around $V=8eV$, it saturates. For comparison, we also show
the charge density $P_e(y)$ in Fig.2 and Fig.3. Here $P_e(y)>0$ in
any position $y$, and $P_e(y)$ is symmetric while $P_s(y)$ is
asymmetric.

\begin{figure}%[tbp]
\includegraphics[bb=10mm 10mm 188mm 131mm, width=8.5cm,totalheight=6cm,clip=]{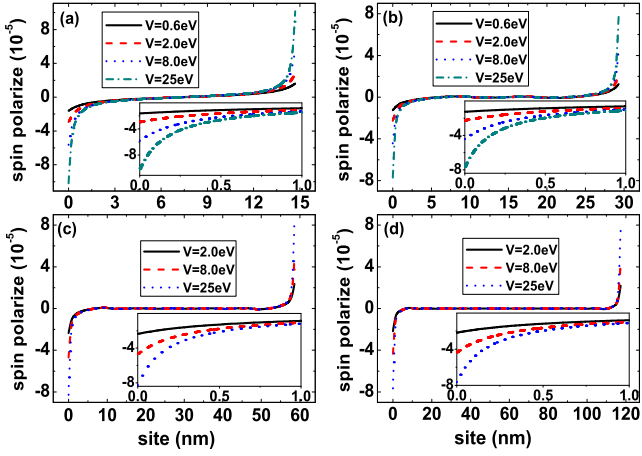}
\caption{ (Color online) The spin polarization $P_s/P_e$ ${\sl
vs}$ the transverse site $y$ for the square confining potential.
Panel (a),(b),(c),(d) correspond to $L=14.6nm$, $L=29.3nm$,
$L=58.6nm$, and $L=117.3nm$, respectively. The insets show the
detailed distribution near the edge of $y=0$ in the central panel.
}
\end{figure}
\begin{figure}%[tbp]
\includegraphics[bb=10mm 9mm 188mm 132mm, width=8.5cm,totalheight=6cm,clip=]{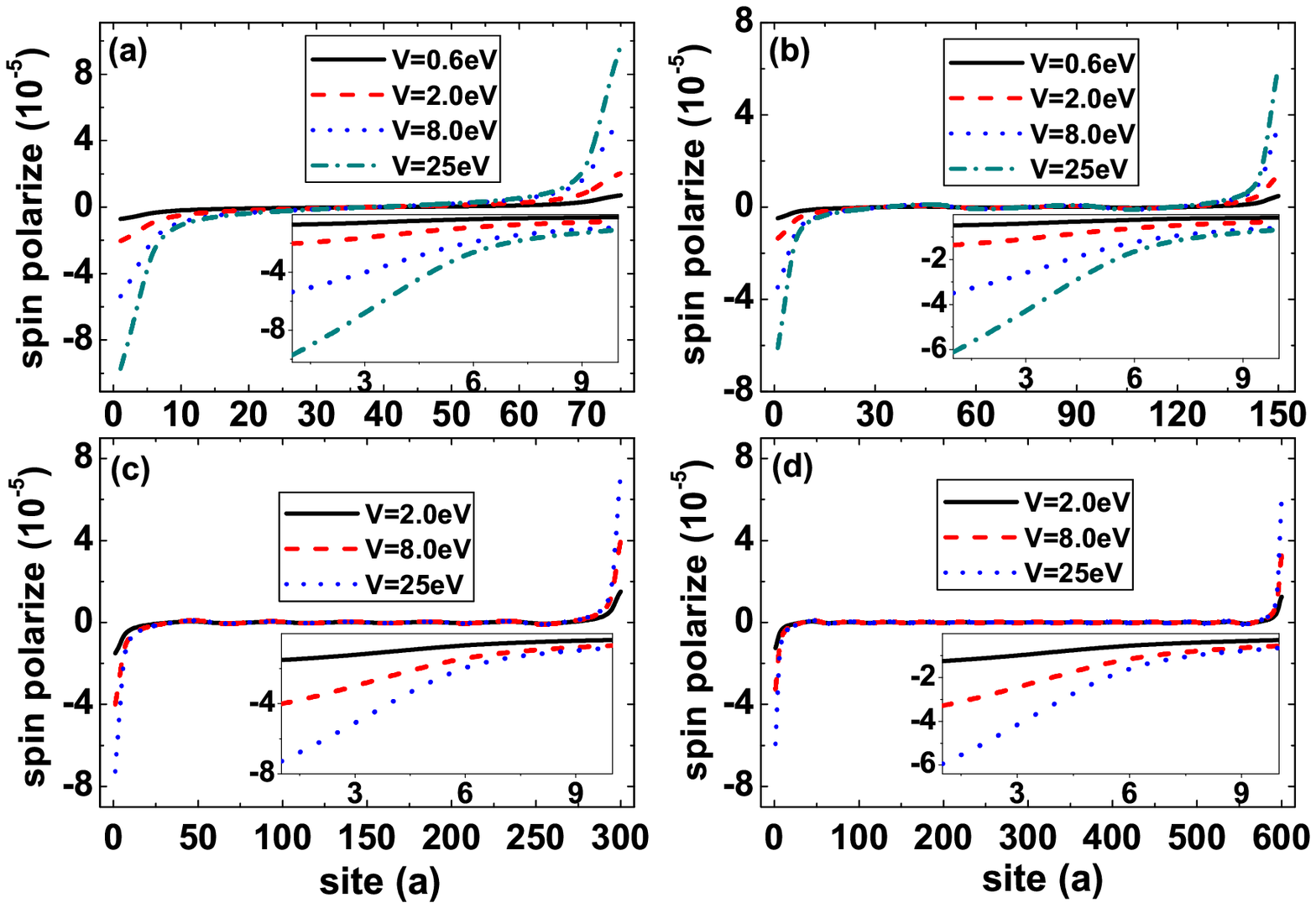}
\caption{ (Color online) The spin polarization $P_s/P_e$ ${\sl
vs}$ the transverse site $y$ for the parabolic confining
potential. Panel (a),(b),(c),(d) correspond to $L=75 a$, $L=150
a$, $L=300 a$, and $L=600 a$, respectively. The insets show the
detailed distribution near the edge of $y=0$ in the central panel.
}
\end{figure}

Next, we discuss the spin polarization $P_s(y)/P_e(y)$. In the
central panel of Fig.4 and Fig.5, we plot the transverse
distribution of the spin polarization for the square and parabolic
confining potential $V(y)$, respectively. It is clearly shown that
the opposite spin polarization emerges on the transverse two edges
whatever $L$ is set. For example, in the Fig.4 and 5, the transverse
width $L$ is set to $14.6nm$, $29.3nm$, $58.6nm$ and $117.3nm$ which
is very different, but the spin polarization distributions near the
edge are almost identical. In addition, the spin polarization on the
transverse edge is  close to a linear function as $V$ while $P_s(y)$
is hardly affected by the strength of $V$ (see Fig.2 and 3). The
detailed distributions of the spin polarization near the edge of $y=0$
are magnified and shown in the insets (in Fig.4 and 5). The
distribution range for the parabolic potential is slightly wider
than that for the square potential because the variation range of
the parabolic potential $V(y)$ is wider than that of the square
potential $V(y)$.

The above result is obtained in a clean system. In the following, we
study the spin accumulation in the dirty system. Fig.6 displays the
transverse distribution of $P_s(y)$, $P_e(y)$ and spin polarization
$P_s(y)/P_e(y)$ with different disorder strength
$W=0.1E_f,1E_f,2E_f.5E_f,10E_f$. In these calculation, $P_s(y)$ and
$P_e(y)$ are obtained by averaging over up to 5000 realizations of
disorder. From Fig.6, we can see that the spin accumulation $P_s(y)$
and the spin polarization near the two edges are almost  unaffected
by the disorder, even the disorder $W$ is very strong (e.g.
$W=10E_f$). In fact, the spin accumulation in the present device
originates from the confining potential near the edge, where the
effective potentials for spin-up and spin-down electrons are
different (see Fig.1). Intuitively, the spin polarization is
expected to be unaffected by the disorder as well as the dephase.
Additionally, with the increasing of the disorder $W$, the
amplitudes of the oscillation of $P_e(y)$ and $P_s(y)$ in the bulk
are slightly reduced (see Fig.6a,b), and their fluctuations are
increased linearly.

To compare our numerical results with the experiment
\cite{experiment1}, we calculate the value for spin polarization and
accumulation. From the experimental data,   the spin density at the
peak can be estimated:  $P_s\approx (1.5--4.2)\times
10^{-6}nm^{-2}eV^{-1}$, and the spin polarization is
$(1.0--2.8)\times 10^{-4}$ for a thickness $h=0.9\mu m$. From
figures (e.g. fig.2 and fig.4),
our calculation show that the peak spin density
$P_s\approx 0.3 \times 10^{-6}nm^{-2}eV^{-1}$ and the spin
polarization is $0.8\times 10^{-4}$. The value of the spin
polarization is comparable with the experiment. Due to the
big thickness in the expriment, the spin accumulation $P_s$
in our calculation seems to be several times smaller than that of the
experiment. Indeed, a more precise quantitative calculation perhaps requires to
consider the spin orbit coupling in the bulk and spin relaxation
effect at the boundary. However, our simple model indeed produces
the same order of magnitude as one measured in the experiments.

%Fig.6
\begin{figure}%[tbp]
\includegraphics[bb=10mm 10mm 206mm 148mm, width=8.5cm,totalheight=5.5cm,clip=]{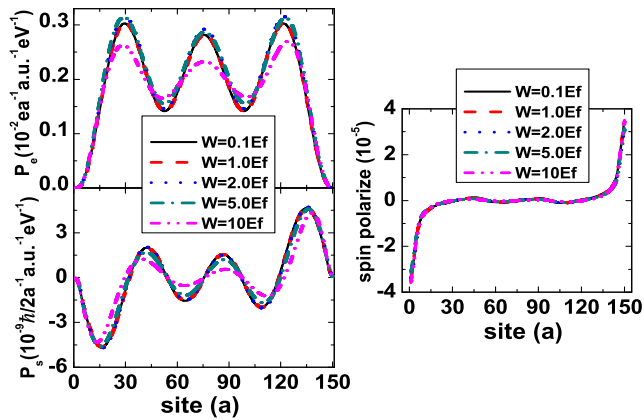}
\caption{(Color online) The charge density distribution $P_e$ (a),
the spin density distribution $P_s$ (b), and spin polarization
$P_s/P_e$ (c) ${\sl vs}$ the transverse site $y$ for the different
disorder $W$, respectively. The other parameters are: $E_f$=0.1eV,
$L=150 a$, and $V=8eV$. }
\end{figure}

\section{Conclusion}

In summary, we propose a new mechanism to explain the spin
accumulations at the edges of a zonal two dimensional electron
system. Due to the strong structure confining potential in the
boundary, the induced  spin-orbit interaction leads to   the
opposite spin accumulation on the two transverse  edges under the
longitudinal voltage bias. The spin polarized direction can be
reversed while the bias is reversed.  The intensity of the
polarization is also proportional to the external longitudinal
voltage bias.  These results are consistent with the recent
experiment.    Moreover, the experimental test of the new mechanism
can be easily performed in future experiments. Unlike the extrinsic
and intrinsic SHE,    the spin accumulations in the present
mechanism are hardly affected by the disorder and dephase, and  can
exist even in the strong disorder system.
\section*{Acknowledgements}

We gratefully acknowledge financial support from the Chinese Academy
of Sciences and NSF-China under Grant Nos. 90303016, 10474125, and
10525418. L. Tang and J.P Hu are also supported by NSF with award
number: PHY-0603759.

\end{document}